\documentstyle[12pt]{article}
\newlength{\pubnumber} 

\catcode`\@=11
\@addtoreset{equation}{section}

\def\section{\@startsection{section}{1}{\z@}{3.5ex plus 1ex minus .2ex}
 {2.3ex plus .2ex}{\large\bf}}
\def\subsection{\@startsection{subsection}{2}{\z@}{2.3ex plus .2ex}
 {2.3ex plus .2ex}{\bf}}

\begin{document}

\begin{titlepage}
\samepage{
\setcounter{page}{1}
\rightline{McGill/92-42}
\rightline{\tt hep-th/9210064}
\rightline{September 1992}
\vfill
\begin{center}
 {\Large \bf Fractional Superstrings with Critical\\
Spacetime Dimensions Four and Six:\\
A Status Report\footnote{Talk presented at the {\it International
Workshop on String Theory, Quantum Gravity, and the Unification
of the Fundamental Interactions}, held in Rome, Italy, 21-26 September 1992;
to appear in Proceedings published by World Scientific.}\\}
\vfill
 {\large Keith R. Dienes\footnote{E-mail address:
dien@hep.physics.mcgill.ca}\\}
\vspace{.25in}
 {\it  Department of Physics\\
McGill University\\
3600 University St.\\
Montr\'eal, Qu\'ebec~H3A-2T8~~Canada\\}
\end{center}
\vfill
\begin{abstract}
 {\rm We provide a short non-technical survey
of the fundamental issues involved in the recently-proposed fractional
superstring theories.  After introducing the basic ideas which underlie these
new string theories, we review their early successes and
outline some related outstanding questions.}
\end{abstract}
\vfill}
\end{titlepage}

\setcounter{footnote}{0}

\hyphenation{pa-ra-fer-mion pa-ra-fer-mion-ic pa-ra-fer-mions }
\hyphenation{su-per-string frac-tion-ally su-per-re-pa-ra-met-ri-za-tion}
\hyphenation{su-per-sym-met-ric frac-tion-ally-su-per}
\hyphenation{space-time-super-sym-met-ric fer-mi-on}
\hyphenation{mod-u-lar mod-u-lar--in-var-i-ant}

\def\ie{{\it i.e.}}
\def\eg{{\it e.g.}}
\def\etc{{\it etc}}
\def\beq{\begin{equation}}
\def\eeq{\end{equation}}
\def\beqn{\begin{eqnarray}}
\def\eeqn{\end{eqnarray}}
\def\bZ{{\bf Z}}
\def\bone{{\bf 1}}
\def\Kac{{Ka\v{c}}}
\def\thetatwo{{\vartheta_2}}
\def\thetathree{{\vartheta_3}}
\def\thetafour{{\vartheta_4}}
\def\ket#1{{|{#1}\rangle}}
\def\calZ{{\cal Z}}
\def\half{{\textstyle{1\over 2}}}
\def\tautwo{{\tau_2}}

\section{Fractional Superstrings:  The Basic Ideas}

Fractional superstrings are
non-trivial generalizations of ordinary superstrings and heterotic strings,
and have critical spacetime dimensions which are less than ten.
In this paper, I aim to provide a self-contained introduction to these
new string theories, giving an overview of the entire field as it currently
stands rather than focusing exclusively on any particular aspect.
The first section of this paper, therefore, will be entirely non-technical,
and will summarize the basic idea behind the fractional superstrings
as well as their early successes and some related outstanding issues.
The second section will then provide more details concerning the construction
of the fractional-superstring worldsheet theories, their partition functions,
and their various possible spacetime interpretations.

\subsection{Motivation:  Why new string theories?}

There have been, historically, a variety of different classes of string
theories which have been constructed and analyzed.  The {\it bosonic string}\/
is characterized by a conformal symmetry on the worldsheet, and contains
worldsheet fields which are the coordinate bosons $X^\mu$.
The critical dimension of this theory, however, is 26, and the theory
suffers from spacetime tachyons and the absence of spacetime fermions.
The {\it superstring}\/ augments the conformal symmetry on the worldsheet
by adding a worldsheet supersymmetry (resulting in a superconformal symmetry),
and has worldsheet fields which are the bosons $X^\mu$ and their
fermionic superpartners $\psi^\mu$.  While this
removes the spacetime tachyons, gives rise to spacetime fermions,
and causes the critical dimension to fall to 10,
it is not clear how to incorporate gauge groups into this theory
in a natural manner or obtain a reasonable spacetime phenomenology.
The (closed) {\it heterotic string}, on the other hand, contains
a superconformal worldsheet symmetry for the right-movers and a
conformal symmetry for the left-movers.  Although the critical dimension
of this theory remains 10, gauge groups appear in a natural manner and
indeed realistic low-energy phenomenologies can emerge.  Thus the
heterotic string has proven to be a fruitful starting point in
string-theoretic model-building.
Even though the critical spacetime
dimension of the heterotic string is 10,
this can be overcome by either {\it compactifying}\/ the
six undesired dimensions (\eg, on a variety of different orbifolds
or Calabi-Yau manifolds) or {\it replacing}\/
them with alternative but ``equivalent'' conformal field theories
(\eg, those involving a sufficient number of free bosons or fermions
to achieve conformal anomaly cancellation).
\smallskip

The problem with this dimensional-reduction ``solution,''
from a phenomenological point of view, is that there are {\it many}\/
different ways of passing from the critical dimension $D_c=10$ to $D=4$,
and each of these thus gives rise to a different model-construction procedure
yielding an entirely new class of models.
This great freedom in constructing models would not in itself
be a problem, of course, were it not for the fact that each resulting
model has a unique
spacetime phenomenology, with its own set of particles, gauge
groups, and interactions.  Thus, the physical properties of the
resulting theory depend crucially on the particular model chosen.
Even this would not be a problem, however, were it not for the observation
that the original choice is completely arbitrary:  string theory,
as presently formulated, leaves all of these choices equally likely,
classically degenerate in the space of possible string field theory
vacua.  While this degeneracy might hopefully be lifted by
string-field-theoretic effects, such a string field theory has yet to
be formulated.
\smallskip

Our goal, then, is clear.  Since the source of this problem was
the great freedom in passing from the critical dimension to the
actual spacetime dimension, we would like to construct new types of
string theories  which have {\it critical}\/ spacetime dimension
$D_c=4$.  We would then have no need for compactification, have
fewer extra degrees of freedom to arbitrarily choose or adjust,
and might possibly be able to construct a truly
unique string theory directly in four dimensions.
There are also, however, less phenomenological reasons for attempting
to construct such theories.  String theories are thought to
embody a symmetry principle that is a generalization of
the usual field-theory notion of gauge invariance to include
higher-spin fields and non-linear interactions.
Our understanding of these spacetime string symmetries remains
sketchy, however, since their connection to structures and symmetries
on the worldsheet is as yet somewhat mysterious.
Another motivation, then, for searching for new such string
theories is to possibly shed light on this connection between
worldsheet and spacetime physics.

\subsection{The Fractional Superstring}

How might we proceed to construct theories with smaller
critical dimensions?  We already have some indications:  we saw
that if there is {\it no} worldsheet supersymmetry (as in the
bosonic string), the critical dimension is $D_c=26$, whereas if we add
a single worldsheet supersymmetry (as in the super- or heterotic string),
the critical dimension falls to $D_c=10$.
A natural guess, then, is to increase the symmetry on the worldsheet.
However, one finds that putting an {\it additional} supersymmetry
on the worldsheet yields a theory with $D_c=2$,\footnote{
These are actually two {\it complex} dimensions, implying a
4D spacetime with $(2,2)$ signature.}
whereas the addition of extra worldsheet supersymmetries turns out
not to further change the critical dimension.
Thus, it is clear that we need something intermediate --
a {\it fractional}\/ supersymmetry.
It turns out that such fractional supersymmetries
can indeed be formulated on the two-dimensional worldsheet,
and the strings based on such worldsheet symmetries are
called ``fractional superstrings''.
\smallskip

We can thus add a new type of string to the
previous list:  the fractional superstring.  The worldsheet symmetry
of this theory is therefore the usual conformal symmetry
plus a fractional supersymmetry, together forming
what is called a fractional superconformal symmetry.
Correspondingly, the worldsheet fields in
this theory consist not only of the
bosons $X^\mu$, but also their {\it fractional}\/ superpartners,
the {\it para}\/fermions $\epsilon^\mu$.  These parafermionic
fields have fractional spin on the worldsheet,
but we stress that in two dimensions such fields are allowed;
in particular,
the presence of such fields on the worldsheet does not necessarily
imply the appearance of fractional-spin particles in spacetime.
Constructing such theories, then, we find that
their critical spacetime dimensions
are indeed less than ten, and the precise value
depends on the fraction chosen for the supersymmetry.
In particular, it turns out that we can parametrize
this fraction by an integer $K\geq 2$ (the degenerate case $K=1$
corresponds to the absence of any supersymmetry).
The special case $K=2$ corresponds to ordinary supersymmetry, which, we recall,
transforms a worldsheet boson $X$ of spin zero into a worldsheet fermion
$\psi$ of spin $\Delta(\psi)=1/2$ via the action of a supercurrent of the
form $J\sim \psi \partial X$ with spin $1+\Delta(\psi)=3/2$.
The cases with $K>2$ then correspond to our fractional
supersymmetries:  for each such $K$, these symmetries transform
a worldsheet boson $X$ of spin zero into
a so-called ``$\bZ_K$ parafermion'' $\epsilon$ of
spin $\Delta(\epsilon)=2/(K+2)$ via the action of a worldsheet fractional
supercurrent of the form $J\sim \epsilon \partial X +...$ with
spin $1+\Delta(\epsilon)$.
Given this parametrization, then,
one finds the corresponding critical spacetime dimensions:
\beq
    D_c ~=~  \cases{
      26 & for $K=1$ \cr
     2 + 16/K & for $K\geq 2$.}
\label{kcritdimensions}
\eeq
We therefore have the following cases with integer values of $K$ and $D_c$:
\beqn
       K=1:&~~~~~~~&D_c=26 \nonumber \\
       K=2:&~~~~~~~&D_c=10,~~{\rm spin}(\psi)=1/2 \nonumber \\
       K=4:&~~~~~~~&D_c=\phantom{1}6,~~{\rm spin}(\epsilon)=1/3 \nonumber \\
       K=8:&~~~~~~~&D_c=\phantom{1}4,~~{\rm spin}(\epsilon)=1/5 \nonumber \\
      K=16:&~~~~~~~&D_c=\phantom{1}3,~~{\rm spin}(\epsilon)=1/9~.
\label{kcases}
\eeqn

Given these fractional supersymmetries, it is possible to generate
many new classes of fractional superstring theories with various critical
spacetime dimensions.  Since we are considering only closed strings,
the left- and right-moving symmetries need not be the same, and thus
we can label our $(K_L,K_R)$ theories according to the fractions of
their left- and right-moving supersymmetries.  In this new language,
we see that the ordinary bosonic string is simply the $(1,1)$ special case of
the fractional superstring,
the ordinary superstring (\eg, the Type II string)
is simply the $(2,2)$ special case,
and the ordinary heterotic string can be identified simply as
the $(1,2)$ string.
The more general $(K_L,K_R)$ strings, however, will have critical spacetime
dimensions less than ten.
In particular, the $(K,K)$ strings have the same fractional supersymmetry
for the left- and right-moving sectors, and are the natural generalizations
of the closed superstring.
Similarly, the $(1,K)$ strings are the natural
generalizations of the heterotic string.

\subsection{Overview of Results and Open Questions}

Since the original fractional-superstring proposal \cite{AT},
there has been much progress in attempting to demonstrate
the self-consistency of these new string theories and
in understanding much of the new physics they contain [2--7].
We shall defer a detailed examination of some of these issues to
Sect.~2, and instead now proceed to give a quick overview
of the results achieved as well as a brief discussion of the many
issues which remain to be completely resolved.
\smallskip

There have been a number of early successes
which indicate that the fractional superstring is likely to be a
consistent theory both on the worldsheet and in spacetime.
First, of course, this approach provides a natural generalization
of the more traditional string theories, establishing
(via the use of two-dimensional fractional supersymmetries and parafermions)
a uniform framework for examining a variety of different string
theories \cite{AT,DT}.
Secondly, as mentioned, these strings have critical spacetime dimensions
which are less than ten, and in particular the phenomenologically
interesting dimensions $D_c=4,6$ are allowed values.  Third, the
low-lying spacetime spectra of these theories have been examined in some
detail \cite{ALT},
and indeed for each relevant value of $K$ one finds a tachyon-free spectrum
containing a graviton, gauge bosons, and massless spacetime fermions;  indeed,
these latter spacetime spinorial states satisfy the Dirac equation by
virtue of a parafermion zero-mode Clifford algebra, in complete analogy
to what occurs in the ordinary superstring.  Fourth, it appears that
these new string theories have a {\it spacetime} supersymmetry \cite{ADT}:
this symmetry appears naturally in these theories, evidenced
by equal numbers of spacetime bosonic and fermionic states
at all mass levels, and is consistent with the appearance
of massless spin-3/2 particles (gravitinos).
Fifth, it has been possible to construct modular-invariant partition
functions for these theories which are consistent with each of the
above properties \cite{AT,DT}, and these in turn have been used profitably
to obtain much insight into some of the more difficult and unique
aspects of these theories \cite{ADT,AD,CR}.
Sixth, promising first steps towards fractional-superstring {\it
model-building}
have been taken \cite{DT}, and as expected one finds that the
set of $K>2$ models is indeed orders of magnitude smaller than the analogous
set of $K=2$ models.  These successes are all, therefore,
signs that the fractional superstring theories will ultimately
prove both consistent and fruitful.
In fact, work in this area has even led to results which are independent
of fractional superstrings {\it per se}, such as whole classes of
new identities relating the characters of $\bZ_K$ parafermions for
different values of $K$ \cite{ADT,AD,FL}.
\smallskip

There remain, however, many different aspects of the fractional
superstrings which still need to be investigated.  While some of these
will be necessary for proving the ultimate consistency of these theories,
others will be important for uncovering the intrinsically
new physics these theories contain.  These different issues manifest
themselves in the worldsheet physics, in the spacetime physics, and in
the mapping between the two.
\smallskip

On the worldsheet, for example, it will be necessary to have a
more complete understanding of the two-dimensional fractional
superconformal symmetries and algebras which are the fundamental
worldsheet building blocks of the fractional superstrings.
Since these algebras involve fields which have fractional spin,
the field-operator product expansions which comprise these
algebras contain {\it cuts} rather than simple
poles, and thus these algebras are necessarily non-local (as well
as non-linear).  Most of the difficulty in dealing with
these algebras stems from this simple fact.
For example, one important issue which has yet to be resolved
is the fractional superghost system -- one will need to understand
the ghosts arising from fractional superconformal gauge-fixing in order
to ``properly'' derive the critical dimension $D_c$ from conformal anomaly
cancellation and thereby demonstrate the consistency of this aspect of
the fractional superstring.
This non-linearity of the fractionally-supersymmetric
boson plus parafermion $(X,\epsilon)$ theory
may also be the key to
many of the intrinsically {\it new} features of these strings.
For example, it is tacitly assumed in the fractional superstring
construction that one can start with a single
such $(X,\epsilon)$ theory corresponding to one given
spacetime dimension, and then simply tensor
together $D_c$ copies of this theory to produce the full $D_c$-dimensional
worldsheet theory;  certainly this would be the straightforward
generalization of the bosonic and superstring constructions.
However, each individual $(X,\epsilon)$ theory
is non-linear and non-local, while a tensor product
is necessarily a linear, local combination (\ie, parafermions
corresponding to different dimensions in the tensor product
do not ``feel'' each other in a non-local way, while those corresponding
to the same dimension do).  Whether such
a straightforward tensor-product construction can produce a fully
self-consistent Lorentz-invariant
theory remains to be seen \cite{ALT}.
\smallskip

A similar situation exists for the spacetime physics of the
fractional superstring.  Here too there are a variety of issues
which still must be examined in order to prove the physical consistency
of the fractional superstring (\eg, one must develop formalisms
for calculating scattering amplitudes and demonstrating unitarity,
as well as, eventually, for {\it bona-fide}\/ model-building and
examining the resulting spacetime phenomenologies \cite{DT,FL}), but once again
there are intrinsically new features which appear and require
closer analysis.  One of these issues is spacetime Lorentz invariance:
while it appears that certain ``sectors'' of the theory are fully
consistent with Lorentz invariance, other new Planck-scale sectors
may be invariant only under {\it subgroups} of the Lorentz
group \cite{DT,ADT}.
This issue is undoubtedly the spacetime reflection of the
worldsheet tensor-product issue discussed above, and whether
the appearance of such new sectors indicates a Planck-scale violation of
Lorentz
invariance (which in itself is not ruled out by experiment) or a highly
non-trivial self-induced spacetime compactification remains to be seen.
These additional sectors also appear to contain Planck-scale
particles which seem to violate the
spin-statistics connection \cite{DT,ADT}.
Properly interpreting these states and their roles in spacetime physics
will be an important endeavor.
\smallskip

Finally, there are various outstanding issues
concerning the {\it mapping}\/ between the worldsheet theory
and the spacetime physics.
In order to map between these different spaces, one
starts with a given worldsheet conformal-field-theory vacuum and
examines the entire Fock space of excitations
on that vacuum;  those states in the Fock space satisfying
certain physical-state conditions are then interpreted
as particles in spacetime.
For the fractional superstrings, it is still an open question
as to what the complete set of physical-state conditions are:
although a subset of these conditions are known \cite{ALT}.
there are many indications that further conditions
having no analogues in the ordinary string theories remain to be found.
It will also then be necessary to develop a no-ghost theorem for the
fractional-superstring Fock space -- \ie, to demonstrate that the
all of the states in the resulting space have non-negative norm.
Additionally, there are various ``projections'' which occur in the
fractional-superstring Fock space and which need to be more
adequately understood.  While some of these projections are analogous
to the ordinary GSO projection which removes tachyonic states from
the spacetime spectrum of the superstring, other so-called ``internal
projections'' have no analogue in the superstring and turn out to be
responsible for removing exponentially large numbers of states
from the theory (thereby changing its Hagedorn
temperature and effective central charge) \cite{ADT,ALT,AD}.
\smallskip

Thus, while the fractional superstring has seen many successes,
it also provides an arena in which many qualitatively new issues can and
must be explored.  Work in all of these areas is actively
being pursued.

\section{Fractional Superstrings:  Some Details}

In order to be more specific concerning some of these outstanding
issues, we must first examine these fractional superstring theories
in more detail.   In this section, therefore, we shall provide a short
tutorial in the fractional-superstring fundamentals, examining the
construction of the fractional-superstring worldsheet theories, their
corresponding partition functions, and the resulting spacetime physics.

\subsection{Constructing the Worldsheet Theory}

Before discussing the worldsheet theory of the fractional superstring,
it proves instructive to re-examine that
of the ordinary superstring (which is the $K=2$ special case).
The worldsheet fields of the superstring are of course the free spin-0
bosons $X^\mu$ and their superpartners, the free Majorana spin-1/2
fermions $\psi^\mu$.  Here $\mu=0,1,...,D_c-1$, and we therefore
have $D_c$ copies of the fundamental supersymmetric $(X,\psi)$ theory, one
copy corresponding to each spacetime dimension.  In order to fully
specify this theory, however, we must choose for our fields any
of four different boundary conditions around the two non-contractible
cycles (spacelike and timelike, respectively) of the torus.
The bosons $X^\mu$ must always have periodic/periodic boundary conditions
around these two cycles (in keeping with their interpretation as spacetime
coordinates), but the fermions $\psi^\mu$, which always appear in
physical quantities through $\psi\psi$
bilinears, may have periodic or antiperiodic boundary conditions.
Choosing $(AP)$ or $(AA)$ boundary conditions for the fermions puts us in the
Neveu-Schwarz sector of the theory, and all excitations in this
sector correspond to spacetime bosons.
Choosing $(PP)$ or $(PA)$ boundary conditions, however, puts us in the
Ramond sector of the theory, and all excitations in this
sector correspond to spacetime fermions.
\smallskip

It turns out that this free Majorana fermion theory which appears on
the superstring worldsheet
can be rewritten in a variety of ways which are useful
for later generalization.   One way is to think of this theory
as the Ising model with central charge $c=1/2$.  The fields in the
Ising model are $\lbrace \bone,\psi,\sigma,\sigma^\dagger\rbrace$,
where $\bone$ is the identity with highest weight $h=0$,
$\psi$ is the fermion with $h=1/2$,
and $\sigma$ and $\sigma^\dagger$ are the spin
field and its conjugate with $h=1/16$.  These four fields
are, of course, linear combinations of the previous Majorana fermion with
its four possible boundary conditions:  indeed, the $\bone$ and $\psi$ fields
combine to yield the Neveu-Schwarz sector, whereas the spin fields $\sigma$
and $\sigma^\dagger$ combine to yield the Ramond sector.  A second
useful way to think of the Majorana-fermion theory
is as the \Kac-Moody coset theory $SU(2)_2/U(1)$:   here the $SU(2)$ is
taken at \Kac-Moody level 2
(as indicated by the subscript), and after modding out by
$U(1)$ the resulting quotient theory is what can be called
a ``$\bZ_2$ parafermion theory''.  We shall describe such quotient
theories in more detail in the
next subsection, but the basic idea is that the fields $\phi^j_m$
in such a theory
fall into representations labeled by quantum numbers $j$ and $m$
inherited from the underlying $SU(2)$ structure.
For this particular theory at level 2, there are only the
four fundamental fields
$\lbrace \phi^0_0, \phi^1_0, \phi^{1/2}_{\pm 1/2}\rbrace$, and these are again
in direct one-to-one correspondence with the four fields of the Ising model,
with $\phi^0_0$ and $\phi^1_0$ identified as the $\bone$ and $\psi$ fields
respectively, and $\phi^{1/2}_{\pm 1/2}$ identified as the spin fields.
Since the contribution to the worldsheet supercurrent from the fields
corresponding to any one spacetime dimension in the superstring is
of course $J=\psi\partial X$, we may rewrite this in
terms of the quotient theory as $J=\phi^1_0 \partial X$.
\smallskip

Now that we have written the superstring ($K=2$) worldsheet theory in
this manner,
it is straightforward to generalize to higher values of $K$.
To construct the worldsheet theory of the fractional superstring
for general $K\geq 2$, we begin with a single copy
of the {\it fractionally}\/-supersymmetric $(X,\phi)$ theory, where
$\phi$ now denotes the so-called ``$\bZ_K$ parafermion theory''
 -- \ie, the \Kac-Moody coset theory $SU(2)_K/U(1)$,
which will be discussed below.  We then tensor together $D_c$
copies of this theory, obtaining $(X^\mu, \phi^\mu)$ where the
spacetime index $\mu=0,1,...,D_c-1$ is understood to
be contracted with the Minkowski metric.
Similarly, our worldsheet {\it fractional}\/ supercurrent per
dimension then takes the analogous form $J\sim \phi^1_0 \partial X+\ldots$,
where $\phi^1_0$ is now a field in the general $\bZ_K$ parafermion theory.
\smallskip

\subsection{$\bZ_K$ Parafermion Theories:  Basic Facts}

These $\bZ_K$ parafermion theories,
although complicated, can be easily summarized by a few simple facts.
For any $K$, the $\bZ_K$ parafermion theories are conformal field theories
with central charges $c=(2K-2)/(K+2)$,
and their primary fields $\phi^j_m$ can be labeled and organized
by their $SU(2)$ quantum numbers $j$ and $m$, where $|m|\leq j$, $j-m\in \bZ$,
and where $j\geq 0$.  One new feature, however, is the fact that $j$
is also bounded from above, $j\leq K/2$, and thus these theories contain
only a finite number of fundamental fields, this number growing with
increasing $K$.  The spins of these fields have an expected
$j(j+1)$ dependence, as is usual for $SU(2)$, but once again there is a
novelty, an $m$-dependence:
\beq
       \Delta^j_m~=~ {{j(j+1)}\over{K+2}} ~-~ {{m^2}\over{K}} ~.
\label{paraspins}
\eeq
The fusion rules of these fields also resemble those of an $SU(2)$
tensor product:
\beq
   \lbrack \phi^{j_1}_{m_1}\rbrack ~\otimes~
   \lbrack \phi^{j_2}_{m_2}\rbrack ~=~
   \sum_{J=|j_1-j_2|}^{J_{\rm max}} \,
   \lbrack \phi^{J}_{m_1+m_2}\rbrack ~,
\label{parafusions}
\eeq
except that here $J_{\rm max}$ is not merely $j_1+j_2$ but rather the
minimum of this quantity and $K-j_1-j_2$.
Indeed, this general new feature, this ``reflection'' symmetry
$j\leftrightarrow K/2-j$, allows us to consistently identify the
fields $\phi^j_m=\phi^j_{m+K}=\phi^{K/2-j}_{-(K/2-m)}$ for values
outside the range $|m|\leq j$.
\smallskip

For any value of $K$, there are certain fields in the $\bZ_K$ parafermion
theory which will play a special role in the fractional superstring.
The $\phi^0_0$ field, as we have seen, is the identity, and will produce
the Neveu-Schwarz vacuum state.  The $\phi^{K/4}_{\pm K/4}$ fields,
by contrast, are the analogues of the $K=2$ spin fields,
and they will correspondingly produce the Ramond vacuum state.
A third special field, as we have seen, is the $\phi^1_0$ field,
often simply denoted $\epsilon$.  This is the energy operator in the
$\bZ_K$ theories, and is the field which appears in the supercurrent and
through which Fock-space excitations take place.
\smallskip

These parafermion theories are of course more complicated than sketched here.
In particular, as we have stated, the spins $\Delta^j_m$ are
not simple half-integers, and thus
the OPEs between these fields will involve branch {\it cuts} rather than
simple poles.  These
theories are hence non-local on the two-dimensional worldsheet.
This also implies that the parafermion mode algebras will not involve
simple commutation or anticommutation relations, but rather {\it generalized}\/
commutation relations.  A discussion of these issues in the context of
fractional superstrings, particularly for the $K=4$ case, can be found in [4].
What makes the $K=4$ case particularly tractable, it turns out, is the
equivalence of the $\bZ_4$ parafermion theory to that of a free boson
compactified on a circle of radius $R=\sqrt{6}$.
Furthermore, although these theories are non-local,
the $K=4$ theory is {\it abelianly braided}\/ (\ie, involves only
one type of cut in the OPEs).  The $K=8$ and $K=16$ cases, by contrast,
do not share this property.

\subsection{The Critical Dimension}

Our discussion of the construction of the fractional-superstring worldsheet
theory indicated that we start with a single fractionally-supersymmetric
$(X,\phi)$ theory (where $\phi$ indicates the $\bZ_K$ parafermion theory),
and then tensor together $D_c$ copies to form the full worldsheet theory.
It is necessary, however, to determine the critical dimensions $D_c$ for each
value of $K$ -- \ie, to derive the result quoted Eq.~(\ref{kcritdimensions}).
The standard method for obtaining this would normally be to
calculate the contribution to the conformal anomaly from the ghosts
arising from fractional-superconformal gauge fixing, and then choose
$D_c$ so that $D_c c_0$ cancels this quantity [where $c_0=3K/(K+2)$ is the
total central charge of each $(X,\phi)$ theory].  The problem, however, is that
the fractional superghost system is not yet understood, and the central
charge of this system has not yet been calculated.  We therefore choose
an alternative path, demanding that our resulting theory indeed be a
string theory including gravity and hence containing a massless graviton.
\smallskip

This argument, which is a straightforward generalization of an old
argument due to Brink and Nielsen, proceeds in two steps.  First we
calculate the vacuum energy (or intercept $-v$) of the theory.
Since we know that conformal invariance requires that
$v=c_{\rm eff}/24$ where $c_{\rm eff}$
is the effective central charge of {\it propagating}\/ degrees of freedom
in the theory, and since (as usual) only the $D_c-2$ transverse
dimensions worth of states yield physically propagating states,
we see that $c_{\rm eff}=(D_c-2)c_0$.  Thus in general we have the
intercept $v=(D_c-2)c_0/24$.  Our second step is to identify
the graviton as a state excited from this vacuum, and then adjust $v$
(and thereby $D_c$) so that this excited state is massless.
For example, in the $K=1$ bosonic
string, the graviton is $X^\mu_{-1}|p\rangle_R
\otimes \tilde{X}^\nu_{-1}|p\rangle_L$ where $\ket{p}_{L,R}$ indicate these
left- and right-moving vacuum states and where the lowest excitation mode of
the $X$-field increases the energy by one unit (note that the
spinless $X^\mu$-field must have integer moding).
Thus the resulting graviton state will be massless if $v=1$, and since
$c_0=1$ for this theory, we obtain $D_c=26$.  For the $K=2$ superstring,
the situation is similar.  Here the
graviton state is $b^\mu_{-1/2}\ket{p}_R\otimes \tilde{b}^\nu_{-1/2}\ket{p}_L$
where the excitation operators are now modes of $\psi$;
since $\psi$ has spin-1/2, it has half-integer moding
and the lowest excitation adds a half-unit of energy.
Thus the graviton will be massless if $v=1/2$, and since $c_0=3/2$
for the $(X,\psi)$ theory, we obtain $D_c=10$.
\smallskip

The fractional-superstring case is analogous.  Here the graviton
state is given by $\epsilon^\mu_{-2/(K+2)}\ket{p}_R\otimes
\tilde{\epsilon}^\nu_{-2/(K+2)}\ket{p}_L$, where once again the moding
of the $\epsilon\equiv \phi^1_0$ field follows from its spin.  This
state will thus be massless if $v=2/(K+2)$, and since $c_0=3K/(K+2)$,
we obtain the result quoted in Eq.~(\ref{kcritdimensions}).  Note that
this set of critical dimensions also guarantees the
existence of {\it massless}\/ spacetime vectors (\ie, gauge bosons)
in the spectra of heterotic $(1,K)$ fractional superstrings,
and also assures us that the chiral light-cone gauge
Ramond vacuum states
$\prod_{\mu=1}^{D_c-2} (\phi^{K/4}_{K/4})^\mu\ket{p}_{L,R}$
will be massless as well [since $(D_c-2)\Delta^{K/4}_{K/4}-v=0$].
This implies that these theories will in general contain
a massless {\it gravitino} state, whereupon consistency requires that they
exhibit full spacetime supersymmetry.  We will see below that this
is indeed the case.

\subsection{Partition Functions}

One method of efficiently surveying the {\it entire}\/ spectrum of states
(rather than merely those that are massless)
is to examine the one-loop partition functions $\calZ(\tau)$
that these theories can have;  indeed, $\calZ(\tau)$ receives
contributions from states at all mass levels:
 ~$\calZ\equiv {\rm Tr} (-1)^F q^H \overline{q}^{\tilde{H}}$
where $\tau$ is the torus modular parameter,
$q\equiv e^{2\pi i \tau}$, and the trace is over the complete Fock
space of states with (left,right) energies ($H$,$\tilde{H}$)
and fermion number $F$.
In order to calculate such partition functions, we must
multiply the characters corresponding to each of the worldsheet fields.
The character of each worldsheet boson
is given by $(\sqrt{\tau_2} \eta\overline{\eta})^{-1}$ where $\tau_2\equiv
 {\rm Im}\,\tau$ and where $\eta\equiv q^{1/24}\prod_{n=1}^\infty
(1-q^n)$ is the Dedekind $\eta$-function;
likewise, the character of each complex (left-moving) worldsheet fermion
generally takes the form $\vartheta/\eta$ where $\vartheta$ represents
one of the four classical Jacobi $\vartheta_i$-functions, depending
on the boundary conditions of the fermion.  Similarly, the character
of each (left-moving) {\it para}\/fermion
$\phi^j_m$ is given by $\eta c^{2j}_{2m}$, where
the $c$'s denote the level-$K$ {\it string functions};
these functions are the higher-$K$ generalizations of the Jacobi
$\vartheta$-functions, and reduce to them in the $K=2$ special case.
(Note that the super- and sub-scripts on the $c$'s are defined
as {\it twice}\/ those on the corresponding $\phi^j_m$.)
Since the worldsheet field content of the $(K,K)$ fractional
superstrings consists (in light-cone gauge) of $D_c-2$ copies
of the $(X,\phi)$ theory, their partition functions must
take the forms $\calZ_K={\tau_2}^k \sum (\overline{c})^{D_c-2} (c)^{D_c-2}$
where $k=1-D_c/2$.  Here the antiholomorphic components come from
the right-moving components of the worldsheet fields,
and the $\eta$-functions have cancelled between the boson
and parafermion characters.
\smallskip

Any linear combinations of the form $\sum (c)^{D_c-2}$
which can appear in these partition functions $\calZ_K$ must satisfy several
conditions.  First, if we $q$-expand these combinations as $\sum a_n q^n$,
we must have $a_n=0$ for all $n<0$ (absence of physical tachyons) as well
as $a_0\not=0$ for some {\it subset}\/ of terms within the linear
combinations (corresponding to a massless sector).  We must also demand
these combinations close under the $S$ modular transformation,
and that our total partition functions $\calZ_K$ be modular invariant.
Remarkably, there exists only a single unique solution satisfying
all of these constraints for each value $K=2,4,8$, and $16$.
These solutions are
\beqn
  \calZ_2&=& \tautwo^{-4}\,\phantom{\lbrace}|A_2|^2 \nonumber\\
  \calZ_4&=& \tautwo^{-2}\,\left\lbrace|A_4|^2 ~+~ 3\, |B_4|^2\right\rbrace
    \nonumber\\
  \calZ_8&=& \tautwo^{-1}\,\left\lbrace|A_8|^2 ~+~ |B_8|^2 ~+~2\,|C_8|^2
       \right\rbrace\nonumber\\
  \calZ_{16}&=& \tautwo^{-1/2}\,\left\lbrace |A_{16}|^2 ~+~ 4\,|C_{16}|^2
   \right\rbrace
\label{partfuncts}
\eeqn
where
\beqn
   A_2&=&\lbrack 8(c^0_0)^7(c^2_0) +56(c^0_0)^5(c^2_0)^3
      +56(c^0_0)^3(c^2_0)^5 +8(c^0_0)(c^2_0)^7 \rbrack~-~
      \lbrack 8(c^1_1)^8 \rbrack ~~\sim q^0 \nonumber\\
  A_4 &= &\lbrack 4(c^0_0+c^4_0)^3(c^2_0)-4(c^2_0)^4\rbrack
	~-~\lbrack 4(c^2_2)^4 -32(c^2_2)(c^4_2)^3\rbrack
            ~~\sim q^0\nonumber\\
  B_4 &= & \lbrack 8(c^0_0+c^4_0)^2(c^2_2)(c^4_2)
    -4(c^2_0)^2(c^2_2)^2  \rbrack -
   \lbrack 4(c^2_0)^2(c^2_2)^2 -16(c^0_0+c^4_0)(c^2_0)(c^4_2)^2
    \rbrack \,\sim q^{1/2}\nonumber\\
  A_8 &= &\lbrack 2(c^0_0+c^8_0)(c^2_0+c^6_0)-2(c^4_0)^2\rbrack
   ~ -~\lbrack 2(c^4_4)^2-8(c^6_4c^8_4)\rbrack ~~\sim q^0\nonumber\\
  B_8 &=&
    \lbrack4(c^0_0+c^8_0)(c^6_4) -2(c^4_0c^4_4)\rbrack ~-~
   \lbrack 2(c^4_0c^4_4) -4(c^2_0+c^6_0)(c^8_4) \rbrack
     ~~\sim q^{1/2}\nonumber\\
  C_8 &= &4(c^2_2+c^6_2)(c^0_2+c^8_2)-4(c^4_2)^2 ~~\sim q^{3/4}\nonumber\\
  A_{16}&= &\lbrack c^2_0+c^{14}_0-c^8_0\rbrack
	  ~-~\lbrack c^8_8-2c^{14}_8\rbrack ~~\sim q^0\nonumber\\
  C_{16}&= &c^2_4+c^{14}_4-c^8_4 ~~\sim q^{3/4}~.
\label{ABC}
\eeqn

Certain features of these partition functions can be easily interpreted.
The $A$-terms, whose expansions begin with $q^0$, contain the contributions
from the massless states in these theories, and indeed we find
that each of these $A_K$ expressions contains a term $+(D_c-2)(c^0_0)^{D_c-3}
(c^2_0)$ corresponding to the massless vector state $\epsilon^\mu\ket{p}$
(recall that $\epsilon\equiv \phi^1_0$).
Similarly, each $A_K$ expression also contains a term
$-(D_c-2)(c^{K/2}_{K/2})^{D_c-2}$ corresponding to the massless fermion state
(the Ramond vacuum);  the minus sign indicates the spacetime
statistics of this state, and once again the coefficient indicates the number
of degrees of freedom appropriate for such spinors.
In fact, it is straightforward to identify those terms within each $A_K$
which correspond to spacetime bosonic or fermionic states
of {\it any}\/ mass.
Since the bosonic Neveu-Schwarz vacuum state (respectively the fermionic
Ramond vacuum state) has its $m$-quantum number equal to zero (respectively
$K/4$) for each of the $D_c-2$ transverse directions,
and since all excitations on these vacua take place through the $\phi^1_0$
field with vanishing $m$-quantum number, we see from the
fusion rules (\ref{parafusions}) that {\it any}\/ excited states must
have the same $m$-quantum numbers as their underlying vacua.
Indeed, since $m$-quantum numbers are additive
modulo $K/2$ under fusion, this yields the usual boson/fermion fusions
$B\otimes B=F$, $B\otimes F=F$, and $F\otimes F=B$.
These observations thus enable us to split
each expression $A_K$ into its separate bosonic and fermionic components,
with the first square-bracketed piece within each $A_K$
in Eq.~(\ref{ABC}) arising from spacetime bosons, and the second
piece (preceded by an overall minus sign) arising from spacetime fermions.
\smallskip

Another feature concerns spacetime supersymmetry.
For $K=2$, we can re-express $A_2$ in terms of the
equivalent $\vartheta$-functions to find that $A_2=\half \eta^{-12}J$
where $J\equiv \thetathree^4-\thetatwo^4-\thetafour^4$;
thus $\calZ_2$ is indeed recognized as the partition function of the ordinary
superstring, and the Jacobi identity $J=0$ (or $A_2=0$)
indicates the vanishing
of $\calZ_2$ (\ie, the exact level-by-level cancellation of bosonic
and fermionic states).  This is of course the partition-function
reflection of the spacetime supersymmetry of this theory.
Remarkably, however, this property extends to higher $K$ as well,
for it can be proven \cite{ADT} that {\it each}\/ of the combinations listed
in Eq.~(\ref{ABC}) vanishes identically: ~$A_K=B_K=C_K=0$.
These resulting new identities, which are the higher-$K$ generalizations
of the $K=2$ Jacobi identity $A_2=0$,
can therefore be taken as evidence of spacetime
supersymmetry in the {\it fractional}\/ superstrings.

\subsection{Open Questions:  New Vacua and Internal Projections}

There are, however, a number of issues concerning these partition
functions which are yet to be resolved and which are undoubtedly
at the heart of the intrinsically new physics of the
fractional superstrings.  These are, essentially, the questions
of new vacuum states and of ``internal projections.''
\smallskip

While it was straightforward to identify each term within the $A_K$-sectors
as arising from either a Neveu-Schwarz or Ramond vacuum state,
it is clear that this is not the case for the $B_K$ or $C_K$ sectors.
Of course, these sectors have $q$-expansions beginning only at
$q^h$ with $h>0$, and thus they do not contribute to massless
(\ie, observable) physics except through loop effects suppressed
by powers of the Planck mass.  Nevertheless, although such sectors do
not appear in the ordinary $K=2$ superstring case,
they do appear as fundamentally new features of the
fractional superstrings, and their roles
(and spacetime-physics properties) are not
yet understood.  The states in the $B_K$-sectors, for example,
appear to originate from ``mixed'' Neveu-Schwarz/Ramond vacuum states
in which {\it half}\/ of the transverse dimensions contribute Neveu-Schwarz
vacua and the other half Ramond:
$(\phi^0_0 \phi^{K/4}_{K/4})^{(D_c-2)/2} \ket{p}$.
The states in the $C_K$-sectors, by contrast,
appear to originate from vacuum states
$(\phi^{K/8}_{K/8})^{D_c-2} \ket{p}$
which are ``pure'' (\ie, the
same in each transverse dimension), but of fractional spin
(because the fusion of {\it two}\/ such vacuum states seems
necessary to produce the Ramond vacuum).
One possible interpretation of the roles of these
new Planck-scale states is as follows \cite{DT}.
Although the states in the $B_K$-sectors appear to violate full
$D_c$-dimensional Lorentz invariance (because their underlying vacua
are not invariant under the permutations of all of the $D_c-2$
transverse spacetime directions), they may well be invariant
under $D$-dimensional {\it subgroups} of the Lorentz group, with $D<D_c$.
The appearance of such states would then provide an intrinsically new
mechanism for Lorentz symmetry breaking --- \ie, for
spontaneous spacetime self-compactification.
Indeed, in the $K=4$ case, the form of $B_4$ suggests
that in order for each corresponding state to be
interpretable as arising from an underlying Neveu-Schwarz or Ramond vacuum,
the resulting spacetime dimension $D$ can be at most four
(\ie, with only {\it two} transverse dimensions);
the two remaining string-function factors would
then correspond to the internal CFT induced by the compactification.
Similarly, in the $K=8$ case, we find that $D$ can be at most three,
in which case the appearance of the fractional-spin $C_8$ sector
would be fully consistent with the spin-statistics connection.
[Note that the $K=16$ theory, which is already has $D_c=3$, requires
(and receives) no additional compactification from any $B$-type sector,
and the presence of a $C_{16}$ sector is already non-problematic.]
Furthermore, it turns out that such self-induced compactifications would
naturally provide a mechanism for the appearance
of phenomenologically necessary chiral spacetime fermions
in the fractional-superstring spectra \cite{DT}.
Of course, detailed analyses are necessary in order to verify these ideas.
\smallskip

A more concrete new issue, perhaps, is the question of the ``internal
projections'' which arise in the $K>2$ fractional superstrings and
which are also reflected in the above partition functions.
We have already seen that we could split each expression
$A_K$ into separate (though cancelling) bosonic and fermionic
components $A_K=A_K^b-A_K^f$, and while the relative minus sign {\it between}
$A_K^b$ and $A_K^f$ is of course attributed to
spacetime statistics [\ie, to the $(-1)^F$ factor in the
definition of $\calZ$],
the minus signs appearing {\it within}
each of these components for $K>2$ clearly cannot be understood as
arising due to spacetime statistics.   Rather, they must be interpreted
as fundamentally new ``internal'' projections which, like the GSO
projection, act between states of the same spacetime statistics.
One might of course worry that in the individual component $q$-expansions
$A_K^{b,f}=\sum a_n q^n$, these extra minus signs might cause some
coefficients $a_n$ to become negative (thereby ensuring that $A_K^b$
and $A_K^f$ could not be the interpreted as the characters of
bosonic or fermionic highest-weight sectors of an underlying
post-projection conformal field theory).  However, it has been
proven \cite{ADT} that {\it each}\/ of
these components individually satisfies the remarkable identity
\beq
      A_K^{f} ~=~ (D_c-2)\,\left(\prod_{n=1}^\infty\,
      {{1+q^n}\over{1-q^n}}\right)^{D_c-2} ~=~
  (D_c-2)\,\left[ {{\thetatwo(\tau)}\over{2\eta^3(\tau)}} \right]^{(D_c-2)/2}~.
\label{Asplit}
\eeq
This result thus not only guarantees that the coefficients $a_n$
are all non-negative, but in fact indicates that the
residual conformal field theories which underlie the
$A_K^f$ sectors of the fractional
superstrings {\it after the internal projections}
are (in light-cone gauge) nothing but those of $D_c-2$ free bosons and
Ramond fermions,
with apparently no trace remaining of the original {\it para}\/fermions!
This implies that the internal projection reduces
the $A$-sector theory of the fractional superstrings to that
of the ordinary Green-Schwarz string formulated in $D_c=3,4,6,$ or $10$
spacetime dimensions.
Now while these are of course precisely the
dimensions in which the {\it classical}\/ Green-Schwarz
string can be formulated, it is well-known that
this string is quantum-mechanically self-consistent only in $10$ dimensions.
Presumably, then, it is the role of the new fractional-superstring
$B$-sectors --- which appear only for $K>2$ (or $D_c<10$) ---
to restore this self-consistency and cancel the quantum anomalies.
It turns out that there exist arguments \cite{AD,CR} indicating
that proper splittings for the $B_K$-sectors are those indicated by
the square-brackets within the $B_K$-expressions in Eq.~(\ref{ABC}),
and indeed one finds \cite{AD} a similar remarkable identity
for the $B$-sectors:
\beq
       B_K^{f}~=~ (D_c-2)\, \left[\,
     {{\thetatwo(\lambda\tau)}\over{2\eta^3(\tau)}}
     \,\right]^{(D_c-2)/2}
\label{Bsplit}
\eeq
where $\lambda = [\Delta(\epsilon)]^{-1}=\half(K+2)$.  This implies
that the internally-projected $B$-sectors of the fractional superstrings
also resemble Green-Schwarz strings in $D_c<10$,
except with fermions formulated on  {\it rescaled}\/ momentum lattices;
indeed, the underlying parafermionic spins
now appear as the relevant scaling factors.
The $C$-sectors, by contrast, appear to be entirely removed by the
internal projections, and thus may contain no physical states of
any spacetime statistics \cite{AD,CR}.
\smallskip

The effects of these internal projections,
as indicated by Eqs.~(\ref{Asplit}) and (\ref{Bsplit}),
are quite profound.
What had begun (in light-cone gauge) as $D_c-2$ copies of a
fractionally-supersymmetric boson plus $\bZ_K$ parafermion conformal field
theory with total central charge $(D_c-2) c_0= 48/(K+2)$
has now been internally projected down to $D_c-2$ copies of a boson
plus (rescaled) fermion theory with total central charge $3(D_c-2)/2=24/K$.
Thus exponentially large numbers of states at each mass level of the
Fock space are being projected out of the spectrum by these internal
projections, enough to reduce the effective central charges
(and thereby alter such features of these theories
as their Hagedorn temperatures,
which depend only on how the numbers of states diverge at infinite mass.)
Indeed, these internal projections indicate that our original parafermionic
Fock space is too large, and that although the fractional superstrings
can be formulated in terms of these parafermionic CFTs {\it with
internal projections}, an alternative (and smaller) CFT
might exist which would more efficiently describe the residual
physical states surviving the internal projections.  Restoring
the contributions from longitudinal and timelike directions, we see that
such residual covariant CFTs would have to have total central charges
$c=24/K + 6K/(K+2)$,
and remarkably this is the spectrum of critical central charges
for proposed string theories \cite{ALTtwo}
based upon spin-$(K+4)/(K+2)$ algebras (\eg, the so-called ``spin-4/3
algebra'' for $K=4$, which has no linear tensor-product formulation).
An important goal for the fractional-superstring theories, then,
is to construct suitable representations of these algebras
with precisely this spectrum of critical central charges.

\bigskip
\medskip
\leftline{\large\bf Acknowledgments}
\medskip

It is a pleasure to thank Philip Argyres and Henry Tye, with whom
I collaborated on much of the work described herein, for many
useful discussions.  I would also like to acknowledge the
hospitality of the Aspen Center for Physics, where portions of
this work were completed.  This work was supported in part
by the Natural Sciences and Engineering Research Council of Canada
and by les Fonds FCAR du Qu\'ebec.

\vfill\eject

\bigskip
\bibliographystyle{unsrt}

\smallskip
\noindent {\bf \underbar{Note}:}~~ Listed above
are only those references concerning
fractional superstrings;  papers on related issues are referenced within
these.

\end{document}